\documentstyle[12pt,axodraw]{article}
\begin{document}
\baselineskip 0.6cm

\def\simgt{\mathrel{\lower2.5pt\vbox{\lineskip=0pt\baselineskip=0pt
           \hbox{$>$}\hbox{$\sim$}}}}
\def\simlt{\mathrel{\lower2.5pt\vbox{\lineskip=0pt\baselineskip=0pt
           \hbox{$<$}\hbox{$\sim$}}}}
\def\x{x}

\begin{titlepage}

\begin{flushright}
FERMILAB-Pub-03/284-T
\end{flushright}

\vskip 2.0cm

\begin{center}

{\Large \bf 
Higgsless Theory of Electroweak Symmetry Breaking from Warped Space
}

\vskip 1.0cm

{\large
Yasunori Nomura
}

\vskip 0.4cm

{\it Theoretical Physics Department, Fermi National Accelerator 
     Laboratory, Batavia, IL 60510} \\

\vskip 1.2cm

\abstract{
We study a theory of electroweak symmetry breaking without a Higgs 
boson, recently suggested by Csaki {\it et al}.  The theory is 
formulated in 5D warped space with the gauge bosons and matter 
fields propagating in the bulk. In the 4D dual picture, the theory 
appears as the standard model without a Higgs field, but with an 
extra gauge group $G$ which becomes strong at the TeV scale. The 
strong dynamics of $G$ breaks the electroweak symmetry, giving the 
masses for the $W$ and $Z$ bosons and the quarks and leptons. 
We study corrections in 5D which are logarithmically enhanced 
by the large mass ratio between the Planck and weak scales, and 
show that they do not destroy the structure of the electroweak 
gauge sector at the leading order.  We introduce a new parameter, 
the ratio between the two bulk gauge couplings, into the theory 
and find that it allows us to control the scale of new physics.
We also present a potentially realistic theory accommodating quarks 
and leptons and discuss its implications, including the violation 
of universality in the $W$ and $Z$ boson couplings to matter and 
the spectrum of the Kaluza-Klein excitations of the gauge bosons. 
The theory reproduces many successful features of the standard 
model, although some cancellations may still be needed to satisfy 
constraints from the precision electroweak data.
}

\end{center}
\end{titlepage}

\section{Introduction}
\label{sec:intro}

Despite its extraordinary successes in describing physics below the 
energy scale of $\sim 100~{\rm GeV}$, the standard model of particle 
physics does not address the mechanism of electroweak symmetry 
breaking.  In the standard model the electroweak symmetry is broken by 
a vacuum expectation value (VEV) of a Higgs field, which is driven by 
a non-trivial potential introduced just to trigger the symmetry breaking. 
Moreover, this potential is not stable against radiative corrections 
and is extremely sensitive to any high energy scales, leading to the 
notorious gauge hierarchy problem.  An elegant idea to address this 
problem is to assume that the electroweak symmetry is broken by 
some gauge dynamics whose interaction becomes strong at the TeV 
scale~\cite{Weinberg:gm}.  This idea has been pursued over 25~years, 
but the attempts of constructing realistic theories have encountered 
many obstacles such as the difficulties of obtaining realistic quark 
and lepton masses, suppressing unwanted flavor violation, and complying 
with precision electroweak data.

While the problems listed above depend in principle on the dynamics 
of the strong gauge sector and there may be some theories free from 
these problems, the strong coupling dynamics makes it difficult to 
address these issues quantitatively and construct realistic theories 
in which the agreement with experiments is reliably seen. This is 
particularly problematic because the naive estimate based on an analogy 
with QCD suggests that generic theories of dynamical electroweak symmetry 
breaking are in disagreement with data~\cite{Peskin:1990zt}. There are 
various attempts for overcoming these obstacles, but none seems 
completely satisfactory.

The AdS/CFT duality~\cite{Maldacena:1997re} offers a new possibility 
of addressing these questions.  Suppose the gauge interaction 
responsible for electroweak symmetry breaking has a large number 
of ``colors'' and is almost conformal at energies larger than the 
dynamical scale.  In this case the theory has an equivalent description 
in terms of a weakly coupled 5D theory defined on the truncated 
anti~de-Sitter (AdS) space~\cite{Arkani-Hamed:2000ds}.  The geometry 
of the 5D theory is that of Randall and Sundrum~\cite{Randall:1999ee}, 
and in this picture the hierarchy between the Planck and the electroweak 
scales is understood in terms of the AdS warp factor.  The original 
picture of dynamical electroweak symmetry breaking is then mapped 
to the following situation in 5D.  The electroweak gauge symmetry, 
with the gauge fields propagating in the bulk, is broken at the 
infrared brane by either boundary conditions or a VEV of the 
order of the local cutoff scale. This line of constructing theories 
of dynamical electroweak symmetry breaking has recently been 
considered in~\cite{Csaki:2003zu} with a new ingredient of 
introducing a gauged custodial symmetry in the bulk of a 5D 
theory~\cite{Agashe:2003zs}, which corresponds in the 4D picture 
to imposing a global custodial symmetry on the strong dynamics.%
\footnote{Models of a composite Higgs boson using AdS/CFT have been 
considered in~\cite{Gherghetta:2003wm,Contino:2003ve,Agashe:2003zs}.}
It has been claimed in~\cite{Csaki:2003zu} that the correct masses 
and couplings of the electroweak gauge bosons are reproduced at 
the level of a percent.  In this paper we first study this issue, 
including effects in 5D which are logarithmically enhanced by 
the large mass ratio between the Planck and the weak scales. 
We explicitly show that these corrections do not destroy the structure 
of the electroweak gauge sector obtained in~\cite{Csaki:2003zu} at 
the leading order.

We also extend the analysis of~\cite{Csaki:2003zu} to the case 
where the couplings of the bulk gauge groups take arbitrary values. 
This allows us to incorporate a realistic structure for quarks and 
leptons relatively easily and to make the scale of new physics higher. 
While we do not perform a detailed analysis for the corrections to the 
electroweak observables at a level of a percent, we find that this extra 
freedom for pushing up the scale of new physics allows us to reduce 
the deviations from the standard model.  In view of these, we think 
it is quite interesting if we could construct a theory of electroweak 
symmetry breaking that does not rely on the Higgs mechanism but 
reproduces successful features of the standard model, including 
the fermion mass generation and smallness of flavor changing neutral 
currents (FCNCs). We thus attempt to construct such theories.

We formulate the theory in the 5D AdS space with the gauge group 
$SU(3)_C \times SU(2)_L \times SU(2)_R \times U(1)_X$ in the bulk. 
These gauge groups are broken to $SU(3)_C \times U(1)_{\rm EM}$ by 
large VEVs for scalar fields located on the branes.  Although we 
formally use scalar field VEVs to break gauge groups, the physical 
Higgs bosons decouple when these VEVs are large so that the theory 
does not contain any narrow scalar resonance. To incorporate the 
observed structure for the quark and lepton masses in a simple way, 
we choose representations for quarks and leptons asymmetric under 
the interchange between $SU(2)_L$ and $SU(2)_R$. We find that the 
theory does not give an immediate contradiction with existing 
experimental data for certain parameter region, although some 
amount of tuning may be necessary to satisfy precision electroweak 
constraints.  We also consider the Kaluza-Klein (KK) towers for the 
gauge bosons and find that they are within the reach of the LHC.

The organization of the paper is as follows.  In the next section 
we summarize the theory and framework.  In section~\ref{sec:calc} 
we calculate the masses and couplings of the electroweak gauge 
bosons at the leading order in the warp factor and show that the 
structure of~\cite{Csaki:2003zu} is preserved despite the presence 
of corrections enhanced by a logarithm of the large mass ratio 
between the Planck and TeV scales.  We give a generalized relation 
between the scale of new physics and the masses of $W$ and $Z$, 
which allows us to push up the masses of the resonances in the sector 
of dynamical electroweak symmetry breaking. In section~\ref{sec:model} 
we present a concrete model in the present context.  We discuss fermion 
mass generation and FCNCs arising from the non-universality of fermion 
couplings to the electroweak gauge bosons.  The spectrum for the KK 
towers of the gauge bosons are also discussed.  Conclusions are 
given in section~\ref{sec:concl}.

\section{Theory and Framework}
\label{sec:theory}

Following Refs.~\cite{Agashe:2003zs,Csaki:2003zu} we consider 
an $SU(3)_C \times SU(2)_L \times SU(2)_R \times U(1)_X$ gauge 
theory in 5D warped space.  The extra dimension is compactified 
on $S^1/Z_2$ and the metric is given by
\begin{equation}
  d s^2 \equiv G_{MN} dx^M dx^N 
    = e^{-2k|y|} \eta_{\mu\nu} dx^\mu dx^\nu + dy^2,
\label{eq:metric}
\end{equation}
where $y$ is the coordinate of the fifth dimension and the physical space 
is taken to be $0 \leq y \leq \pi R$; $k$ is the AdS curvature scale. 
We take $k$ to be around the 4D Planck scale, $k \sim M_{\rm pl}$, and 
$kR \sim 10$ to address the gauge hierarchy problem~\cite{Randall:1999ee}. 
The gauge kinetic terms in the bulk are given by
\begin{eqnarray}
  S &=& \int\!d^4x \int\!dy \, \sqrt{-G}
    \Biggl[ -\frac{1}{4g_L^2} g^{MP}g^{NQ} \sum_{a=1}^{3} 
      F^{La}_{MN} F^{La}_{PQ} 
  -\frac{1}{4g_R^2} g^{MP}g^{NQ} \sum_{a=1}^{3} 
      F^{Ra}_{MN} F^{Ra}_{PQ} 
\nonumber\\
  && \qquad \qquad \qquad \qquad
  -\frac{1}{4g_X^2} g^{MP}g^{NQ} F^{X}_{MN} F^{X}_{PQ} \Biggr],
\label{eq:bulk-kinetic}
\end{eqnarray}
where $F^{La}_{MN}$, $F^{Ra}_{MN}$ and $F^{X}_{MN}$ are the 
field-strength tensors for $SU(2)_L$, $SU(2)_R$ and $U(1)_X$, 
and $g_L$, $g_R$ and $g_X$ are the 5D gauge couplings having 
mass dimensions $-1/2$; $a$ is the indices for the adjoint 
representation of $SU(2)$.  Here, we have suppressed $SU(3)_C$, 
since it is irrelevant for our discussion below.

To reduce the gauge symmetry down to $SU(3)_C \times U(1)_{\rm EM}$ 
at low energies, where $U(1)_{\rm EM}$ is the electromagnetism, 
we impose non-trivial boundary conditions on the gauge fields 
at $y=0$ (the Planck brane) and $y=\pi R$ (the TeV brane). They are 
given by~\cite{Csaki:2003zu,Csaki:2003dt}
\begin{equation}
  \partial_y A^{La}_\mu = 0, \quad 
  A^{R1,2}_\mu = 0, \quad
  \partial_y \Bigl( \frac{1}{g_R^2}A^{R3}_\mu 
    + \frac{1}{g_X^2}A^X_\mu \Bigr)=0, \quad
  A^{R3}_\mu - A^X_\mu=0,
\label{eq:bc-0}
\end{equation}
at $y=0$ and 
\begin{equation}
  \partial_y \Bigl( \frac{1}{g_L^2}A^{La}_\mu 
    + \frac{1}{g_R^2}A^{Ra}_\mu \Bigr) = 0, \quad 
  A^{La}_\mu - A^{Ra}_\mu = 0, \quad 
  \partial_y A^X_\mu = 0,
\label{eq:bc-pi}
\end{equation}
at $y=\pi R$; the $A_5$'s obey Dirichlet (Neumann) boundary conditions 
if the corresponding $A_\mu$'s obey Neumann (Dirichlet) boundary 
conditions.  Here, we have retained arbitrary bulk gauge couplings 
$g_L, g_R$ and $g_X$ for later purposes.  In the 4D dual picture, 
this leads to the following situation.  The theory below $k \sim 
M_{\rm pl}$ contains a gauge interaction with the group $G$, whose 
coupling evolves very slowly over a wide energy interval below $k$. 
This $G$ gauge sector possesses a global $SU(3)_C \times SU(2)_L 
\times SU(2)_R \times U(1)_X$ symmetry whose $SU(3)_C \times SU(2)_L 
\times U(1)_Y$ subgroup is weakly gauged, where $U(1)_Y$ is a linear 
combination of $U(1)_X$ and the $T_3$ direction of $SU(2)_R$. 
Therefore, the theory in this energy interval appears as $SU(3)_C 
\times SU(2)_L \times U(1)_Y \times G$ gauge theory (with quarks 
and leptons transforming under $SU(3)_C \times SU(2)_L \times U(1)_Y$; 
see below).  At the TeV scale the gauge interaction of $G$ becomes 
strong. This dynamically breaks $SU(2)_L \times U(1)_Y$ down to 
$U(1)_{\rm EM}$, giving masses to the $W$ and $Z$ bosons (and to 
quarks and leptons).

The boundary conditions of Eqs.~(\ref{eq:bc-0},~\ref{eq:bc-pi}) can 
also be viewed as the limiting case of the following brane Higgs 
breaking.  Suppose that we have a Higgs field $\Sigma({\bf 1}, 
{\bf 1}, {\bf 2}, 1/2)$ on the Planck brane and $H({\bf 1}, {\bf 2}, 
{\bf 2}^*, 0)$ on the TeV brane, where the numbers in the parentheses 
represent gauge quantum numbers under $SU(3)_C \times SU(2)_L \times 
SU(2)_R \times U(1)_X$, and that these fields have VEVs:
\begin{equation}
  \langle \Sigma \rangle = \pmatrix{0 \cr v_\Sigma},
\qquad
  \langle H \rangle = \pmatrix{v_H & 0 \cr 0 & v_H}.
\label{eq:b-higgs}
\end{equation}
For large $v_\Sigma$ and $v_H$ (more precisely $v_\Sigma, v_H \gg 
(k/g_{L,R,X}^2)^{1/2}$) this Higgs breaking resembles the boundary 
condition breaking described above, because a large brane mass term 
for a gauge field pushes off the wavefunction of the gauge field 
from the brane and thus is equivalent to imposing the Dirichlet 
boundary condition~\cite{Nomura:2001mf}. The phenomenology becomes 
identical to that of the boundary condition breaking in the limit 
$v_\Sigma, v_H \rightarrow \infty$ as far as the gauge field sector 
is concerned.  Note that the physical Higgs bosons arising from 
$\Sigma$ and $H$ decouple in this limit and no scalar particle remains 
in the spectrum.  Our analyses below are applicable to both cases for 
the boundary condition breaking and the Higgs breaking described here.

To see whether the theory can be realistic or not, we have to consider 
the couplings of the gauge bosons to quarks and leptons.  Here we 
consider the case of the Higgs breaking but the results should also 
apply to any potentially realistic theories with the boundary condition 
breaking.  A single generation of quarks and leptons corresponds to:
\begin{eqnarray}
  && q({\bf 3}, {\bf 2}, {\bf 1}, 1/6), \qquad
  \bar{u}({\bf 3}^*, {\bf 1}, {\bf 2}, -1/6)|_{T^R_3 = -1/2}, \qquad
  \bar{d}({\bf 3}^*, {\bf 1}, {\bf 2}, -1/6)|_{T^R_3 = 1/2}, 
\nonumber\\
  && l({\bf 1}, {\bf 2}, {\bf 1}, -1/2), \qquad
  \bar{e}({\bf 1}, {\bf 1}, {\bf 2}, 1/2)|_{T^R_3 = 1/2}, \qquad
  [\bar{n}({\bf 1}, {\bf 1}, {\bf 2}, 1/2)|_{T^R_3 = -1/2}],
\label{eq:fermion-rep}
\end{eqnarray}
where $q,\bar{u},\bar{d},l,\bar{e}$ and $\bar{n}$ are left-handed 
fermions and the numbers in the parentheses represent gauge quantum 
numbers under $SU(3)_C \times SU(2)_L \times SU(2)_R \times U(1)_X$;
$T^R_3 = \pm 1/2$ represents the $T_3 = \pm 1/2$ component of the 
$SU(2)_R$ doublet.  A realistic structure for the quark and lepton 
sector is obtained, for example, as follows.  We introduce the quark 
and lepton fields in the bulk, transforming as $q({\bf 3}, {\bf 2}, 
{\bf 1}, 1/6)$, $\psi_{\bar{u}}({\bf 3}^*, {\bf 1}, {\bf 2}, -1/6)$, 
$\psi_{\bar{d}}({\bf 3}^*, {\bf 1}, {\bf 2}, -1/6)$, $l({\bf 1}, 
{\bf 2}, {\bf 1}, -1/2)$ and $\psi_{\bar{e}}({\bf 1}, {\bf 1}, 
{\bf 2}, 1/2)$ [here we do not consider neutrino masses for simplicity]. 
They are Dirac fermions but due to the orbifold compactification 
on $S^1/Z_2$, only the left-handed fermions with the gauge quantum 
numbers given in the parentheses remain massless.  The wavefunction 
profiles for these massless modes are controlled by the bulk mass terms 
for the fermions~\cite{Grossman:1999ra,Gherghetta:2000qt}, which 
we parameterize as ${\cal L}_{\rm 5D} \supset -c k \bar{\Psi} \Psi$ 
where $\Psi$ represents generic 5D (Dirac) fermions.  For $c > 1/2$ 
($c < 1/2$) the wavefunction for the left-handed zero-mode fermion 
is localized to the Planck (TeV) brane.  We take parameters $c$ to 
be larger than $1/2$, at least for the first-two generation fermions. 
The unwanted zero modes, the $T^R_3 = 1/2$ component of $\psi_{\bar{u}}$ 
and the $T^R_3 = -1/2$ components of $\psi_{\bar{d}}$ and 
$\psi_{\bar{e}}$, are made heavy to get masses of order $k$, 
by introducing the Planck-brane localized left-handed fermions 
$\psi'_{\bar{u}}({\bf 3}, {\bf 1}, {\bf 1}, -1/3)$, 
$\psi'_{\bar{d}}({\bf 3}, {\bf 1}, {\bf 1}, 2/3)$ and 
$\psi'_{\bar{e}}({\bf 1}, {\bf 1}, {\bf 1}, 0)$ and couplings 
$\delta(y) [\psi_{\bar{u}} \psi'_{\bar{u}} \Sigma + \psi_{\bar{d}} 
\psi'_{\bar{d}} \Sigma^\dagger + \psi_{\bar{e}} \psi'_{\bar{e}} 
\Sigma^\dagger]$.  The low-energy matter content is then precisely 
that of the standard model quarks and leptons.

The bulk mass terms of $c > 1/2$ are required/beneficial for several 
reasons.  First, since the electroweak gauge symmetry is broken by 
boundary conditions or a large VEV of the brane-localized Higgs, the 
wavefunctions of the $W$ and $Z$ bosons are not flat in the extra 
dimension.  This generically introduces non-universality of the 
electroweak gauge couplings depending on the bulk fermion mass 
parameters, because the 4D gauge couplings are obtained by convolving 
the wavefunctions of the corresponding fermion with the gauge boson, 
which are not universal for fermions having different values of the 
bulk masses.  This non-universality, however, is very small when $c$'s 
are larger than $1/2$ (more detailed discussions are given in 
section~\ref{sec:model}).  Thus we take $c>1/2$ at least for the 
first-two generation fermions [$c$ will be smaller for the right-handed 
top to give a large enough top quark mass].  Second, since fermion 
masses arise from operators localized on the TeV brane [bare 
mass terms in the case of boundary condition breaking, and masses 
through operators like $q \psi_{\bar{u}} H$, $q \psi_{\bar{d}} H$ 
and $l \psi_{\bar{e}} H$ in the case of Higgs breaking], having 
$c > 1/2$ suppresses the masses of the corresponding fermions, 
giving an understanding of the lightness of the first-two 
generation fermions through the localization of the fermion 
wavefunctions~\cite{Gherghetta:2000qt,Huber:2000ie}.  Finally, by 
localizing the wavefunctions to the Planck brane, {\it i.e.} making 
$c$'s larger than $1/2$, we can suppress effects from unwanted 
tree-level flavor-violating operators on the TeV brane.  In any 
event, with $c$ larger than $1/2$, the couplings of the quarks 
and leptons to the electroweak gauge bosons can be effectively 
read off from 
\begin{equation}
  {\cal D}_\mu \psi_L = \Biggl[ \partial_\mu \psi_L + \frac{i}{2} 
    \pmatrix{ A^{L3}_\mu + 2q A^X_\mu & A^{L1}_\mu - i A^{L2}_\mu \cr
    A^{L1}_\mu + i A^{L2}_\mu & -A^{L3}_\mu + 2q A^X_\mu} \psi_L
    \Biggr] \Biggr|_{y=0}
\label{eq:fgc-left}
\end{equation}
for left-handed fermions $\psi_L$ transforming $({\bf 2}, {\bf 1}, q)$ 
under $SU(2)_L \times SU(2)_R \times U(1)_X$, and from 
\begin{equation}
  {\cal D}_\mu \psi_R = \Biggl[ \partial_\mu \psi_R + \frac{i}{2} 
    \pmatrix{ A^{R3}_\mu + 2q A^X_\mu & A^{R1}_\mu - i A^{R2}_\mu \cr
    A^{R1}_\mu + i A^{R2}_\mu & -A^{R3}_\mu + 2q A^X_\mu} \psi_R
    \Biggr] \Biggr|_{y=0}
\label{eq:fgc-right}
\end{equation}
for right-handed fermions $\psi_R$ transforming $({\bf 1}, {\bf 2}, q)$ 
under $SU(2)_L \times SU(2)_R \times U(1)_X$.

It has been shown in~\cite{Csaki:2003zu} that, for fermion couplings 
of Eqs.~(\ref{eq:fgc-left},~\ref{eq:fgc-right}) and the gauge kinetic 
terms given by Eq.~(\ref{eq:bulk-kinetic}) with $g_L = g_R$, the 
tree-level structure of the standard model gauge couplings are 
reproduced.  The relations among various $\gamma$, $W$ and $Z$ 
couplings and the masses of the $W$ and $Z$ bosons are then exactly 
those of the tree-level standard model, up to corrections of order 
$1/\pi kR = O(1\%)$.  Although this level of agreement is not 
sufficient to satisfy constraints from precision electroweak data, 
we think it is interesting if a theory without a Higgs boson 
reproduced the standard model relations at this level. Thus we 
take the attitude that the theory is successful if it reproduced 
the standard model relations at the leading order in $1/\pi kR$. 
In the next section we extend the analysis of~\cite{Csaki:2003zu} 
to the case with logarithmically enhanced corrections in 5D and 
arbitrary bulk gauge couplings.  We explicitly show that the 
successful structure in the electroweak sector is preserved under 
these extensions, and derive a generalized relation between the 
scale of new physics and the masses of $W$ and $Z$.

\section{Masses and Couplings of the Electroweak Gauge Bosons}
\label{sec:calc}

Let us start by considering the 4D dual picture of the theory. In this 
picture, the gauge interaction $G$ becomes strong at the TeV scale, 
producing resonances of masses of order TeV.  These resonances have 
a tower structure.  In particular, there are towers of spin-$1$ fields 
which have the quantum numbers of $W$, $Z$ and $\gamma$.  These towers 
then mix with the elementary gauge bosons of the weakly gauged $SU(2)_L$ 
and $U(1)_Y$ groups.  The resulting spectrum consists of towers of gauge 
bosons with the quantum numbers of $W$ and $Z$, whose lowest states are 
massive and identified as the standard-model $W$ and $Z$ bosons, and 
a tower of $U(1)$ gauge bosons, whose lowest mode is massless and 
identified with the photon.  These towers of mass eigenstates are 
dual to the $W$, $Z$ and $\gamma$ KK towers in the 5D picture.

In the previous section, we have considered the theory in 5D with 
the gauge kinetic terms given by Eq.~(\ref{eq:bulk-kinetic}).  This 
corresponds in the 4D picture to the following situation. In the 
absence of the $G$ sector, the couplings of the elementary gauge 
bosons $g_E$ are very large, $1/g_E^2 \simeq 0$. The observed values 
for the $W$, $Z$ and $\gamma$ couplings, $\alpha = g^2/4\pi$, are 
then obtained purely through the contribution from the $G$ sector. 
In this case, the masses and couplings of $W$, $Z$ and $\gamma$ 
obtained after the elementary-composite mixings respect the 
tree-level standard-model relations at the leading order in $1/N$ 
and $\alpha$, where $N$ is the number of ``colors'' for $G$, since 
this corresponds in the 5D picture to deriving these quantities 
at tree level.  The question is whether the above structure 
is preserved under corrections at higher orders in $1/N$ and 
$\alpha$.  Potentially dangerous ones are loops of elementary fields, 
as they induce non-zero values of $1/g_E^2$ and could change the 
structure.  These corrections are loop suppressed but enhanced by 
a logarithm, so have a size of order $(c \alpha/4\pi)\ln(k/T)$, 
where $c$ is a group theoretical factor of $O(1)$ and $T \equiv 
k e^{-\pi kR} \sim {\rm TeV}$ is the mass scale for the resonances.
Therefore, we want to explicitly see that these corrections do not 
destroy the successful relations among the electroweak gauge boson 
masses and couplings.

In the 4D picture, the relevant corrections of order 
$(c \alpha/4\pi)\ln(k/T)$ arise in the couplings of the elementary 
gauge fields.  In a suitable normalization for the gauge bosons, 
they appear in the coefficients of the gauge kinetic terms for 
the elementary gauge bosons.  In the 5D picture, this effect of 
running gauge couplings for the elementary gauge bosons corresponds 
to the radiatively-induced logarithmic energy dependence for the 
gauge two-point correlators whose end points are both on the Planck 
brane~\cite{Goldberger:2002cz}. This effect is then resummed into 
the coefficients of the Planck-brane localized gauge kinetic terms 
defined at the renormalization scale of order $T$ measured in terms 
of the 4D metric $\eta_{\mu\nu}$. This means that to find the 
corrections to the electroweak observables we simply have to calculate 
the masses and couplings of the $\gamma$, $W$ and $Z$ bosons in the 
presence of Planck-brane localized gauge kinetic terms which have 
coefficients of order $(b/8\pi^2)\ln(k/T)$, where $b$ is the beta 
function coefficient. This corresponds to defining the theory at 
a renormalization scale of order $T$ measured in terms of the 4D 
metric, integrating out the physics above this energy scale.%
\footnote{A related calculation using the renormalization scheme 
adopted here can be found in~\cite{Nomura:2003qb}, where various 
renormalization schemes and their relations are discussed in detail.}

What brane-localized kinetic terms should be included on the 
Planck brane at the 4D renormalization scale of order $T$? 
If the theory possesses bare Planck-brane localized gauge kinetic 
terms, their coefficients can take different values for $SU(2)_L$, 
$SU(2)_R$ and $U(1)_X$, which we denote as $1/\tilde{g}_{L,0}^2$, 
$1/\tilde{g}_{R,0}^2$ and $1/\tilde{g}_{X,0}^2$.  In the 4D picture, 
these couplings correspond to tree-level gauge couplings for the 
elementary $SU(2)_L$, $SU(2)_R$ and $U(1)_X$ gauge fields at a 
scale of order $k$.%
\footnote{In the discussions here and below, we implicitly assume 
the brane Higgs breaking case with large $v_\Sigma$ and $v_H$, 
but essential features of our analysis are unchanged even for 
the boundary condition breaking case.}
On top of this, renormalization group evolution between the energy 
scales of $k$ and $T$ induces an additional contribution to the gauge 
couplings.  Because the elementary gauge group below the scale $k$ 
is $SU(2)_L \times U(1)_Y$, this contribution will produce Planck-brane 
localized kinetic terms for the $SU(2)_L$ and $U(1)_Y$ gauge fields, 
$A^{La}_\mu$ and $A^{R3}_\mu + A^X_\mu$.  We denote coefficients of 
these terms by $1/\tilde{g}_L^2$ and $1/\tilde{g}_Y^2$, which 
take the values $1/\tilde{g}_L^2 = (b_L/8\pi^2)\ln(k/T)$ and 
$1/\tilde{g}_Y^2 = (b_Y/8\pi^2)\ln(k/T)$ where $b_L$ and $b_Y$ 
are the beta function coefficients for $SU(2)_L$ and $U(1)_Y$. 
Adding all together, we obtain the Planck-brane localized kinetic 
terms at the scale $T$:
\begin{eqnarray}
  S &=& \int\!d^4x \int\!dy \, \delta(y) \Biggl[ 
  -\frac{1}{4\tilde{g}_{L,0}^2} 
    \sum_{a=1}^{3} F^{La}_{\mu\nu} F^{La}_{\mu\nu} 
  -\frac{1}{4\tilde{g}_{R,0}^2} 
    \sum_{a=1}^{3} F^{Ra}_{\mu\nu} F^{Ra}_{\mu\nu} 
  -\frac{1}{4\tilde{g}_{X,0}^2} F^{X}_{\mu\nu} F^{X}_{\mu\nu} 
\nonumber\\
  && \qquad \qquad \qquad \qquad
  -\frac{1}{4\tilde{g}_L^2} 
    \sum_{a=1}^{3} F^{La}_{\mu\nu} F^{La}_{\mu\nu}
  -\frac{1}{16\tilde{g}_Y^2} 
    \bigl( F^{R3}_{\mu\nu} + F^X_{\mu\nu} \bigr) 
    \bigl( F^{R3}_{\mu\nu} + F^X_{\mu\nu} \bigr) \Biggr],
\label{eq:brane-kinetic}
\end{eqnarray}
where $F^{R3}_{\mu\nu} \equiv \partial_\mu A^{R3}_\nu - \partial_\nu 
A^{R3}_\mu$.  The point here is that, while the bare couplings 
$1/\tilde{g}_{L,0}^2$, $1/\tilde{g}_{R,0}^2$ and $1/\tilde{g}_{X,0}^2$ 
are parameters of the theory and can take arbitrary values as long as 
they are positive, radiatively induced ones, $1/\tilde{g}_L^2$ 
and $1/\tilde{g}_Y^2$, are calculable quantities and cannot just 
be made smaller.  In fact, the sizes of these radiative pieces are 
$\sim (b/8\pi^2)\ln(k/T) = O(1)$ so we have to include the effect 
of these pieces.  Note that, contrary to the bare couplings 
$1/\tilde{g}_0^2$, radiative couplings can take either negative 
($1/\tilde{g}^2 < 0$) or positive ($1/\tilde{g}^2 > 0$) value, 
depending on whether the corresponding gauge interaction is 
asymptotically free or non-free.  The negative $1/\tilde{g}^2$ is not 
problematic for the radiative piece, because this piece is vanishing 
at the scale $k$ and is generated only at lower scale (in the 4D metric).
At lower energies the gauge couplings squared for physical 4D gauge 
bosons receive contributions both from the brane piece (which could 
be negative) and the bulk piece (which is positive), so that the 
physical gauge couplings squared are always positive.

We now calculate the masses and couplings of the electroweak 
gauge bosons in the theory described in section~\ref{sec:theory}. 
We neglect $SU(3)_C$ in the analysis below, since it is irrelevant 
for our discussion.  For simplicity, we here assume that the 
bare brane kinetic terms are small, {\it i.e.} $1/\tilde{g}_{L,0}^2, 
1/\tilde{g}_{R,0}^2, 1/\tilde{g}_{X,0}^2 \simeq 0$, and include 
only radiative pieces.  In fact, this situation is naturally realized 
by assuming that the theory is strongly coupled at the scale $k$: 
$\tilde{g}_{L,0} \sim \tilde{g}_{R,0} \sim \tilde{g}_{X,0} \sim 4\pi$. 
The effects of introducing bare couplings are discussed at the end 
of this section.

We first solve mass eigenvalues of the $W$ and $Z$ 
gauge-boson towers.  The gauge kinetic terms are given by 
Eqs.~(\ref{eq:bulk-kinetic},~\ref{eq:brane-kinetic}), and 
the boundary conditions for the gauge fields are given by 
Eqs.~(\ref{eq:bc-0},~\ref{eq:bc-pi}) appropriately modified by 
the presence of the brane kinetic terms.  In particular, we derive 
the masses of the lowest towers of $W$ and $Z$, which we identify 
as the standard-model electroweak gauge bosons.  Considering the 
sizes of various couplings, $1/g_L^2 \sim 1/g_R^2 \sim 1/g_X^2 
\sim 1/\pi R$ and $1/\tilde{g}_L^2 \sim 1/\tilde{g}_Y^2 \sim 1$,%
\footnote{Because the 4D gauge couplings, $g_{\rm 4D}$, are 
$O(1)$ and related to the bulk gauge couplings, $g_B$, by 
$1/g_{\rm 4D}^2 \sim \pi R/g_B^2$, $g_B$'s should be regarded 
as quantities of $O(\sqrt{\pi R})$.}
we obtain the following expressions for the $W$ and $Z$ boson masses:
\begin{equation}
  m_W^2 =
    \frac{1}{\frac{\pi R}{g_L^2} + \frac{1}{\tilde{g}_L^2}}
    \frac{2\, T^2}{(g_L^2 + g_R^2)k},
\label{eq:mw}
\end{equation}
\begin{equation}
  m_Z^2 =
    \frac{\frac{\pi R}{g_L^2} + \frac{\pi R}{g_R^2} + \frac{\pi R}{g_X^2} 
      + \frac{1}{\tilde{g}_L^2} + \frac{1}{\tilde{g}_Y^2}}
    {\Bigl(\frac{\pi R}{g_L^2} + \frac{1}{\tilde{g}_L^2}\Bigr)
      \Bigl(\frac{\pi R}{g_R^2} + \frac{\pi R}{g_X^2} 
      + \frac{1}{\tilde{g}_Y^2}\Bigr)}
    \frac{2\, T^2}{(g_L^2 + g_R^2)k},
\label{eq:mz}
\end{equation}
at the leading order in $T/k$ and in $1/\pi kR$.  We have checked 
numerically that these expressions approximate the true values 
at the level of a percent, which is sufficient for our purpose.
The photon mass is zero, $m_\gamma = 0$, as it should be.

The couplings of the electroweak gauge bosons to the quarks and leptons 
can be read off from Eqs.~(\ref{eq:fgc-left},~\ref{eq:fgc-right}).
By normalizing wavefunctions such that the kinetic terms for the 4D 
gauge fields are canonically normalized, {\it i.e.} ${\cal L}_{\rm 4D} 
= - (1/4)F_{\mu\nu}F_{\mu\nu}$, these equations give fermion couplings 
to $\gamma$, $W$ and $Z$ bosons.  We define the fermion gauge coupling 
parameters $g_W$, $e$, $g_Z$ and $\eta_Z$ as
\begin{eqnarray}
  {\cal D}_\mu \psi_L &=& \partial_\mu \psi_L 
    + \frac{i}{2} g_W \pmatrix{ 0 & W^1_\mu-iW^2_\mu \cr
      W^1_\mu+iW^2_\mu & 0 } \psi_L 
\nonumber\\
 && + i e \pmatrix{ Q_{{\rm EM},L\uparrow} & 0 \cr
      0 & Q_{{\rm EM},L\downarrow} } A^\gamma_\mu \psi_L
    + i g_Z \pmatrix{ -Y + \frac{1}{2}\eta_Z & 0 \cr
      0 & -Y - \frac{1}{2}\eta_Z } Z_\mu \psi_L
\label{eq:left-coupling}
\end{eqnarray}
for left-handed fermions $\psi_L = q, l$, and 
\begin{equation}
  {\cal D}_\mu \psi_R = \partial_\mu \psi_R 
    + i e Q_{{\rm EM},R} A^\gamma_\mu \psi_R
    + i g_Z (-Y) Z_\mu \psi_R
\label{eq:right-coupling}
\end{equation}
for right-handed fermions $\psi_R = \bar{u}, \bar{d}, \bar{e}$.
Here, $W_\mu$, $A^\gamma_\mu$ and $Z_\mu$ are the canonically normalized 
4D electroweak gauge bosons, and $Q_{\rm EM}$'s and $Y$ are the electric 
charges and hypercharges.  In the present theory, we find that correct 
values for the electric charges and hypercharges, $Q_{{\rm EM},L\uparrow}$, 
$Q_{{\rm EM},L\downarrow}$, $Q_{{\rm EM},R}$ and $Y$, are reproduced 
for $q$, $\bar{u}$, $\bar{d}$, $l$ and $\bar{e}$ (and $\bar{n}$ 
as well) with the quarks and leptons in the representations of 
Eq.~(\ref{eq:fermion-rep}).  The couplings $g_W$, $e$, $g_Z$ 
and $\eta_Z$ are given by
\begin{eqnarray}
  && \frac{1}{g_W^2} \simeq 
    \frac{\pi R}{g_L^2} + \frac{1}{\tilde{g}_L^2},
\label{eq:gw} \\ \nonumber\\
  && \frac{1}{e^2} \simeq 
    \frac{\pi R}{g_L^2} + \frac{\pi R}{g_R^2} + \frac{\pi R}{g_X^2} 
    + \frac{1}{\tilde{g}_L^2} + \frac{1}{\tilde{g}_Y^2},
\label{eq:e} \\ \nonumber\\
  && \frac{1}{g_Z^2} \simeq 
    \Biggl( \frac{\pi R}{g_R^2} + \frac{\pi R}{g_X^2} 
    + \frac{1}{\tilde{g}_Y^2} \Biggr) 
    + \Biggl( \frac{\pi R}{g_L^2} + \frac{1}{\tilde{g}_L^2} \Biggr)^{-1}
    \Biggl( \frac{\pi R}{g_R^2} + \frac{\pi R}{g_X^2} 
    + \frac{1}{\tilde{g}_Y^2} \Biggr)^2,
\label{eq:etaz} \\ \nonumber\\
  && \frac{1}{\eta_Z} \simeq 
    \Biggl( \frac{\pi R}{g_L^2} + \frac{1}{\tilde{g}_L^2} \Biggr)
    \Biggl( \frac{\pi R}{g_R^2} + \frac{\pi R}{g_X^2} 
    + \frac{1}{\tilde{g}_Y^2} \Biggr)^{-1},
\label{eq:gz}
\end{eqnarray}
at the leading order in $T/k$ and in $1/\pi kR$, and there 
is no other couplings than those of the form given in 
Eqs.~(\ref{eq:left-coupling},~\ref{eq:right-coupling}).

Now, we compare our results, Eqs.~(\ref{eq:mw},~\ref{eq:mz},%
~\ref{eq:gw},~\ref{eq:e},~\ref{eq:etaz},~\ref{eq:gz}), with the 
corresponding tree-level expressions in the standard model.
In the standard model, the masses and couplings of the electroweak 
gauge bosons are given by 
\begin{equation}
  m_W^2 = \frac{g^2}{2} v^2, \qquad
  m_Z^2 = \frac{g^2+g'^2}{2} v^2,
\end{equation}
and
\begin{equation}
  \frac{1}{g_W^2} = \frac{1}{g^2}, \qquad
  \frac{1}{e^2} = \frac{1}{g^2} + \frac{1}{g'^2}, \qquad
  \frac{1}{g_Z^2} = \frac{1}{g'^2} + \frac{g^2}{g'^4}, \qquad
  \frac{1}{\eta_Z} = \frac{g'^2}{g^2},
\end{equation}
respectively, where $g$ and $g'$ represent the $SU(2)_L$ and $U(1)_Y$ 
gauge couplings and $v$ is the VEV of the Higgs field.  From these 
expressions we find that the present theory completely reproduce the 
tree-level structure of the standard model at the level considered here. 
The correspondence between the two theories is given by
\begin{equation}
  \frac{1}{g^2} = \frac{\pi R}{g_L^2} + \frac{1}{\tilde{g}_L^2}, \qquad
  \frac{1}{g'^2} = \frac{\pi R}{g_R^2} + \frac{\pi R}{g_X^2} 
    + \frac{1}{\tilde{g}_Y^2}, \qquad
  v^2 = \frac{4\, T^2}{(g_L^2 + g_R^2)k}.
\label{eq:corresp}
\end{equation}
This generalizes the result of~\cite{Csaki:2003zu} to the case with the 
arbitrary brane kinetic terms and the bulk gauge couplings, especially 
to $g_L \neq g_R$.

Note that the brane pieces $1/\tilde{g}_{L,Y}^2$ cannot just be 
neglected because they are calculable and have size of order 
$(b/8\pi^2)\ln(k/T) \sim 1$. In fact, assuming that the theory has 
the minimal matter content, {\it i.e.} three generations of quarks 
and leptons with the quantum numbers given in Eq.~(\ref{eq:fermion-rep}), 
and unwanted fields are made heavy by Planck-brane masses of order $k$, 
the running of the brane gauge couplings $\tilde{g}_L$ and $\tilde{g}_Y$ 
is given by that of the standard model without a Higgs boson. 
The brane kinetic terms at the scale $T$ are then given by
\begin{equation}
  \frac{1}{\tilde{g}_L^2} =
    \frac{b_L}{8\pi^2}\ln\Bigl(\frac{k}{T}\Bigr)
    \simeq -1.26,
\qquad
  \frac{1}{\tilde{g}_Y^2} =
    \frac{b_Y}{8\pi^2}\ln\Bigl(\frac{k}{T}\Bigr)
    \simeq 2.53,
\label{eq:brane-num}
\end{equation}
where $(b_L,b_Y) = (-10/3,20/3)$ and we have taken $\ln(k/T) 
\simeq 30$.  Therefore, any computation in the present theory 
must include the effect of these terms. 

The effect of bare Planck-brane gauge couplings can be included by 
replacing $\tilde{g}_L^2$ and $\tilde{g}_Y^2$ in Eqs.~(\ref{eq:mw},%
~\ref{eq:mz},~\ref{eq:gw},~\ref{eq:e},~\ref{eq:etaz},~\ref{eq:gz}) as 
\begin{equation}
  \frac{1}{\tilde{g}_L^2} \:\:\rightarrow\:\: 
    \frac{1}{\tilde{g}_L^{\prime 2}}
    = \frac{1}{\tilde{g}_{L,0}^2} + \frac{1}{\tilde{g}_L^2}, 
\qquad\qquad
  \frac{1}{\tilde{g}_Y^2} \:\:\rightarrow\:\: 
    \frac{1}{\tilde{g}_Y^{\prime 2}}
    = \frac{1}{\tilde{g}_{R,0}^2} + \frac{1}{\tilde{g}_{X,0}^2} 
    + \frac{1}{\tilde{g}_Y^2}.
\label{eq:replace}
\end{equation}
Therefore, it does not affect the masses and couplings of the 
electroweak gauge bosons either, at the leading order in $T/k$ and 
$1/\pi kR$.  Incidentally, introduction of bare brane couplings 
does not allow to have $1/\tilde{g}_L^{\prime 2} \simeq 
1/\tilde{g}_Y^{\prime 2} \simeq 0$ in the simplest setup considered 
here, because Eq.~(\ref{eq:brane-num}) and positive bare couplings, 
$1/\tilde{g}_{R,0}^2, 1/\tilde{g}_{X,0}^2 > 0$, imply 
$1/\tilde{g}_Y^{\prime 2} \simgt 2.5$.  Note that, contrary to the 
radiative ones, bare couplings cannot be negative, since they represent 
physical gauge couplings for the elementary gauge bosons at high 
energies so that negative bare couplings lead to the appearance of 
ghosts in the physical states. 

The correspondence obtained in Eq.~(\ref{eq:corresp}) has some 
important physical consequences.  Suppose we numerically calculate 
the deviations from the standard model relations by exactly solving 
the masses and couplings of $W$, $Z$ and $\gamma$.  We then find that 
they are suppressed by $1/\pi kR$ and thus at a level of a percent 
as claimed in~\cite{Csaki:2003zu}.  Moreover, we find that with 
the observed 4D couplings fixed by Eq.~(\ref{eq:corresp}), these 
$1/\pi kR$ deviations become smaller when $g_R/g_L$ becomes larger.
This is because in this parameter region the value of $T$ becomes 
larger and thus the masses of the resonances become higher (see 
the discussion around Eq.~(\ref{eq:KK-mass}) below).  This is an 
important result because it allows us to control the scale of new 
physics in theories with dynamical electroweak symmetry breaking, 
by changing the parameters of the theory.

While we do not attempt to perform a full analysis of higher order 
effects, such as $1/\pi kR$ suppressed corrections and matter loop 
effects, we think it is significant that there is a theory reproducing 
the standard model predictions at a level of a percent or less. 
Because the experimental data require deviations from the standard 
model relations to be smaller than a factor of a few times $10^{-3}$ 
for the electroweak gauge boson masses and couplings, it will still 
be needed to suppress the deviations by a factor of a few to an order 
of magnitude.  These may be attained by pushing up the scale of $T$ 
by changing $g_R/g_L$ and/or by cancellations among various contributions 
such as those from Planck-brane gauge couplings, calculable radiative 
corrections and TeV-brane localized gauge couplings.  Although the 
situation is not completely clear, in the next section we assume that 
these are somehow attained and present a (potentially) realistic theory 
of electroweak symmetry breaking without a Higgs boson.  We find 
that the theory can reproduce many successful features of the standard 
model, including fermion mass generation, while satisfying various 
constraints from experiments.  We also discuss phenomenological 
implications of the theory such as small non-universality of 
fermion couplings to the $W$ and $Z$ bosons and the spectrum 
of the gauge-boson KK resonances.

\section{Potentially Realistic Theory and Its Implications}
\label{sec:model}

In this section we present a theory of electroweak symmetry breaking 
without a Higgs boson, along the line of section~\ref{sec:theory}.
In the 4D dual picture, the theory below the scale $k$ appears 
as the standard model without a Higgs field. It also contains 
an extra gauge group $G$ which becomes strong at the electroweak 
scale, breaking the electroweak symmetry. We present a complete model 
including the quark and lepton sector which reproduces the observed 
structure for the quark and lepton masses and couplings. Some 
elements for constructing our theory can be found in the 
literatures~\cite{Csaki:2003zu,Csaki:2003dt,Agashe:2003zs,%
Gherghetta:2000qt,Huber:2000ie}; in particular, the basic gauge 
structure of the theory has been given in~\cite{Csaki:2003zu}.  
We also discuss several phenomenological implications of the theory.

As already discussed in section~\ref{sec:theory}, we consider an 
$SU(3)_C \times SU(2)_L \times SU(2)_R \times U(1)_X$ gauge theory 
in 5D warped space, with the metric given in Eq.~(\ref{eq:metric}). 
We here consider breaking these groups by Higgs fields localized at 
the Planck and TeV branes, whose representations and VEVs have been 
given in the paragraph containing Eq.~(\ref{eq:b-higgs}).  In the limit 
that $v_\Sigma$ and $v_H$ are large, this breaking becomes equivalent 
to imposing boundary conditions in Eqs.~(\ref{eq:bc-0},~\ref{eq:bc-pi}); 
in particular, the physical Higgs bosons decouple from low-energy 
physics and the theory does not contain any narrow scalar resonance.
At low energies unbroken gauge symmetry of the theory is $SU(3)_C 
\times U(1)_{\rm EM}$, and the weak interaction is mediated by the 
massive $W$ and $Z$ bosons. 

Three generations of quarks and leptons are introduced in the bulk 
as discussed in section~\ref{sec:theory} [in the paragraph containing 
Eq.~(\ref{eq:fermion-rep})].  Specifically, a single generation of 
quarks and leptons are contained in $q({\bf 3}, {\bf 2}, {\bf 1}, 
1/6)$, $\psi_{\bar{u}}({\bf 3}^*, {\bf 1}, {\bf 2}, -1/6)$, 
$\psi_{\bar{d}}({\bf 3}^*, {\bf 1}, {\bf 2}, -1/6)$, $l({\bf 1}, 
{\bf 2}, {\bf 1}, -1/2)$ and $\psi_{\bar{e}}({\bf 1}, {\bf 1}, 
{\bf 2}, 1/2)$ with the unwanted components made heavy by the 
orbifold compactification and mixing with fields localized on 
the Planck brane.  This reproduces the standard model matter content 
at low energies.  The fermion masses arise from the couplings 
introduced on the TeV brane
\begin{equation}
  S = \int\!d^4x \int\!dy \: \delta(y-\pi R) \sqrt{-g_{\rm ind}} 
    \Biggl[ y_u q \psi_{\bar{u}} H + y_d q \psi_{\bar{d}} H 
    + y_e l \psi_{\bar{e}} H + {\rm h.c.} \Biggr],
\end{equation}
where we have suppressed the generation index, $i=1,2,3$.  As the 
up-type quark, down-type quark, and charged lepton masses arise 
from three independent couplings, $y_u$, $y_d$ and $y_e$, it is 
clear that there are no unwanted relations among them coming 
from $SU(2)_R$.  Small neutrino masses can be naturally obtained 
through the seesaw mechanism, by introducing right-handed neutrino 
fields $\psi_{\bar{n}}({\bf 1},{\bf 1},{\bf 2},1/2)$ in the bulk 
together with the TeV-brane couplings $\delta(y-\pi R) y_n 
l \psi_{\bar{n}} H$ and the Planck-brane Majorana masses 
$\delta(y) (\psi_{\bar{n}} \Sigma^\dagger)^2$.%
\footnote{The operators $\delta(y-\pi R) l^2 H^2$ and $\delta(y-\pi R) 
l^2 H^{\dagger 2}$, which could potentially give large masses for 
the observed neutrinos, are forbidden by gauge invariance.}
We choose the bulk masses for the first-two generation fermions 
to be larger than $k/2$, {\it i.e.} $c > 1/2$ [for the definition 
of $c$'s, see section~\ref{sec:theory}].  For $c > 1/2$, the zero 
modes for the fermions have the wavefunctions localized to the 
Planck brane, so that the 4D masses for these quarks and leptons 
are naturally suppressed.  This wavefunction localization also 
suppresses effects from unwanted flavor violating operators on 
the TeV brane, such as $q_i^\dagger q_j q_k^\dagger q_l$.%
\footnote{The amount of localization consistent with the fermion 
masses is not enough to suppress proton decay caused by operators 
on the TeV brane, but this can easily be forbidden by imposing the 
baryon number symmetry on the theory.}
The bulk masses for the third generation $q$ and $\psi_{\bar{d}}$ 
can be chosen to be around $c = 1/2$.  To obtain the top quark 
mass we consider the right-handed top quark to have $c < 1/2$.  Note 
that in our theory the value of $T$ is not strictly related to the 
$W$ and $Z$ boson masses [see Eqs.~(\ref{eq:rel},~\ref{eq:T-example}) 
below], so that we can have larger values of $T$ by choosing 
parameters of the model.  Successful top phenomenology will require 
the value of $T$ close to its maximum, $T \simeq 900~{\rm GeV}$, 
but we leave a detailed study of this issue for future work.  The 
value of $T$ also becomes important when we discuss KK towers 
of the gauge bosons.

We now consider fermion couplings to the electroweak gauge bosons.
As we have seen in the previous section, these couplings are those 
of the standard model for Planck-brane localized matter, at the 
leading order in $1/\pi kR$.  We here assume that this is the case 
at all orders for the Planck-brane matter and consider the effects 
of delocalizing matter in the fifth dimension.  Since the $W$ 
and $Z$ boson wavefunctions have non-trivial profiles in the extra 
dimension, this introduces flavor non-universality in the gauge 
boson couplings to matter and could potentially be dangerous. 
We first consider the $W$ boson couplings to the left-handed fields 
$q$ and $l$.  These couplings are obtained by convolving the 
wavefunction for $q$ (or $l$) with that of the $W$ boson and thus 
depend on the bulk mass for the fermion field. However, we find 
that the resulting non-universality is highly suppressed for $c$ 
larger than $1/2$.  It is less then $10^{-4}$ for $c \simgt 0.6$ and 
less than a percent even for $c = 0.5$. This effect, therefore, does 
not give a contradiction to experimental data.  The delocalization also 
introduces the couplings of the right-handed fields, $\bar{u},\bar{d}$ 
and $\bar{e}$, to $W$, and these couplings have a similar size to the 
non-universality discussed above [$10^{-4}$ for $c \simgt 0.6$ and 
at a level of a percent for $c = 0.5$]. They, however, do not 
connect between the known particles, say $\bar{u}$ and $\bar{d}$, 
but connect the standard-model particles with some heavy fields, 
say $\bar{u}$ with the other component of $\psi_{\bar{u}}$. 
Therefore, this effect is also negligible.

How about the non-universality of the $Z$ couplings?  This is 
potentially dangerous because it induces FCNCs. However, we 
again find that the deviation from the universality is small: 
$\simlt 10^{-4}$ for $c \simgt 0.6$, and even for $c \simeq 1/2$ 
it is less than a percent for left-handed fermions and at most 
a few percents for right-handed fermions. Although this level 
of non-universality could still lead to FCNCs at a level of 
the experimental bounds, here we stick to rough estimates and 
consider that the theory can evade the bounds by having $c$'s 
larger than about $0.6$ for the first-two generation fermions.%
\footnote{We can always choose $c$'s for the first-two generation 
fermions with the same gauge quantum numbers to be equal (or very 
close) to evade the bounds from FCNCs.  The fermion mass hierarchy 
is then generated by the structure of the TeV-brane couplings, 
$y_u$, $y_d$ and $y_e$.}
The effect could be larger for the third generation fields because 
of the smaller values for $c$.  This may have observable consequences, 
but we do not perform a detailed analysis here for these processes.
FCNCs generated through the KK resonances are expected to have a 
similar size.  Incidentally, the fermion couplings to the photon 
are completely universal due to the unbroken $U(1)_{\rm EM}$ gauge 
invariance (because the wavefunction for the photon is flat), so 
no experimental constraint arises from these couplings.%
\footnote{The ratios between the triple gauge-boson vertex 
$\gamma WW$ and the fermion couplings to $\gamma$ are exactly 
those of the standard model for the same reason, while the 
vertex $ZWW$ receives corrections of $O(1\%)$.}

We finally discuss KK towers for the gauge bosons. The masses for 
these towers are given, regardless of the values for the bulk and 
brane gauge couplings, by
\begin{equation}
  m_n \simeq \frac{\pi}{2} \Bigl( n + \frac{1}{2} \Bigr) T,
\end{equation}
where $n=1,2,3,4\cdots$ for the $W$ and $Z$ towers and $n=1,3,5,\cdots$ 
for the $\gamma$ tower (the gluon tower has the same spectrum as the 
$\gamma$ tower).  What is the value of $T$ in the present theory?  
From Eqs.~(\ref{eq:mw},~\ref{eq:mz}) we find that it is related to the 
masses of the $W$ and $Z$ bosons.  This relation, however, does not 
uniquely determine the value of $T$ with the given $W$ and $Z$ masses.
Neglecting small corrections, the relation can be written as
\begin{equation}
  T^2 \simeq 
  \frac{\pi kR}{2 g_W^2} \frac{g_L^2}{\pi R} 
  \left( 1 + \frac{g_R^2}{g_L^2} \right) m_W^2,
\label{eq:rel}
\end{equation}
so we find that $T$ depends on the ratio of the bulk gauge couplings 
$g_R^2/g_L^2$.  The values of $g_L$ and $g_R$, however, are constrained 
to reproduce the observed $W$ and $Z$ boson couplings.  Thus, with the 
given brane kinetic terms, we can determine the allowed region for $T$ 
and consequently the masses for the KK towers. 

As an example, let us consider the case with the minimal matter 
content.  In this case, the brane kinetic terms are given by 
$1/\tilde{g}_L^2 \simeq -1.26$ and $1/\tilde{g}_Y^2 \simeq 2.53$ 
(see Eq.~(\ref{eq:brane-num})),%
\footnote{Having $c < 1/2$ for the right-handed top quark may give 
somewhat smaller values for $1/\tilde{g}_Y^2$ because the value of 
$b_Y$ becomes smaller depending on the value of $c$ for $c < 1/2$.
The resulting correction, however, is small.}
so that the bulk gauge couplings satisfy the relations $\pi R/g_L^2 
= 1/g_W^2 - 1/\tilde{g}_L^2 \simeq 3.61$ and $\pi R/g_R^2 + \pi R/g_X^2 
= 1/e^2 - 1/g_W^2 - 1/\tilde{g}_Y^2 \simeq 5.32$.  Here, we have 
used the experimentally measured values of $g_W \simeq 0.652$ and 
$e \simeq 0.313$.  While $\pi R/g_L^2$ is uniquely determined, 
$\pi R/g_R^2$ is not, although the latter relation implies 
$\pi R/g_R^2 \simlt 5.32$.  The lower bound on the value of 
$\pi R/g_R^2$ then comes from the requirement that the 5D theory 
is not strongly coupled up to the cutoff scale, $M_*$, of the theory. 
This gives $\pi R/g_R^2 \simgt c \pi M_* R/24\pi^3 \approx 0.1 M_*/k$. 
Requiring $M_*/k \simgt 3$ so that the AdS solution is trustable, 
we find $0.7 \simlt g_R^2/g_L^2 \simlt 12$.  Using this in 
Eq.~(\ref{eq:rel}), together with $\pi kR \simeq 30$ and 
$m_W \simeq 80~{\rm GeV}$, we obtain
\begin{equation}
  330~{\rm GeV} \simlt T \simlt 900~{\rm GeV},
\label{eq:T-example}
\end{equation}
which is translated into the values for the masses of the first 
KK towers of $\gamma$, $W$ and $Z$ (and gluon):
\begin{equation}
  800~{\rm GeV} \simlt m_1 \simlt 2.1~{\rm TeV}.
\label{eq:KK-mass}
\end{equation}
Here, smaller (larger) values for $m_1$ correspond to smaller (larger) 
values for $g_R^2/g_L^2$.  Although the couplings of these excitations 
to the matter fields are suppressed by small wavefunction overlaps 
for $c > 1/2$, the above mass region is still within the reach of 
the LHC~\cite{Davoudiasl:2000wi}.  The flexibility for the masses of 
the KK towers also helps to control the size of corrections to the 
electroweak observables.  In fact, for $g_R \gg g_L$, the theory 
does not have any new particle than the standard model gauge bosons 
and quarks and leptons up to an energy scale of $2~{\rm TeV}$, except 
possible states associated with the top quark $t'$.  In particular, no 
physical Higgs boson appears below, and even above, this energy scale.%
\footnote{In spite of the absence of a physical Higgs boson, unitarity 
of the theory is maintained up to a high energy of order the 4D 
Planck scale.  This is clear in the 5D picture because the theory 
is a well-defined effective field theory below the cutoff scale 
$M_*$, which appears close to the 4D Planck scale for an observer 
on the Planck brane.  From the low-energy bottom-up point of 
view, bad high energy behaviors of the longitudinal $WW$ 
scattering are unitarized by the appearance of the KK states 
and the effective vertices arising by integrating out these 
states~\cite{Csaki:2003dt,Csaki:2003zu,SekharChivukula:2001hz}.
This will make the theory to be unitary up to the scale around 
$10~{\rm TeV}$, above which the simple KK picture is not necessarily 
appropriate.  However, the theory above this scale is simply an 
unbroken $SU(3)_C \times SU(2)_L \times U(1)_Y \times G$ gauge 
theory so it is unitary up to very high energies.  In fact, this 
unitarity consideration suggests that the first KK excitation of 
the electroweak gauge bosons appears below $\sim 2~{\rm TeV}$, where 
the $E^2$ amplitude of the longitudinal $WW$ scattering would blow up. 
It is interesting that the maximum value of the lowest KK mass 
obtained in Eq.~(\ref{eq:KK-mass}) saturates this bound.}
It will be interesting to see in detail whether theories such as 
the one described here can really have acceptable parameter regions 
which comply with all the experimental data including the precision 
electroweak data.

\section{Conclusions}
\label{sec:concl}

We have considered a theory of electroweak symmetry breaking without 
a Higgs boson. The theory is formulated in 5D warped space, but 
through AdS/CFT duality, it is related to the 4D theory in which 
the electroweak symmetry is dynamically broken by non-perturbative 
effects of some gauge interaction $G$.  We have studied several issues 
in this framework, including masses and couplings of the electroweak 
gauge bosons and the structure for the quark and lepton mass generation.

We have studied radiative corrections to the masses and couplings of 
the electroweak gauge bosons in 5D which are logarithmically enhanced 
by the mass ratio between the Planck and TeV scales.  We explicitly 
showed that these corrections do not destroy the successful relations 
among the masses and couplings of $W$, $Z$ and $\gamma$ at the leading 
order.  We have also extended the previous result in~\cite{Csaki:2003zu} 
to the case of most general brane and bulk gauge couplings.  This 
allows us to incorporate quarks and leptons in a relatively simple 
manner and also leads to an extra free parameter in the relation 
between the scale of new physics and the $W$ and $Z$ boson masses.
The latter point seems particularly interesting as it allows us to push 
up the masses of the resonances and suppress the deviations from the 
standard model relations arising from loops of these states. Although 
it is still not completely clear whether the theory can naturally 
evade all the precision constraints, it is certainly encouraging 
if we could have a theory without a Higgs boson which successfully 
reproduces the standard model structure at a level of a percent or less. 
We have presented a concrete such model which can accommodate realistic 
quark and lepton masses and mixings as well as their couplings to 
the electroweak gauge bosons.  An interesting possibility is that 
the observed quark and lepton mass structure is obtained through 
wavefunction profiles of these fields in the extra dimension. 
This leads to some amount of flavor non-universality in the matter 
couplings to the gauge bosons. We have estimated this effect and 
found that it does not lead to an immediate contradiction with the 
experimental bounds.  We have also considered the KK excited states for 
the gauge bosons and found that they are within the reach of the LHC.

We have not performed a full analysis for the corrections to the 
electroweak observables at a level of order a percent.  These 
corrections come from tree-level brane kinetic terms at the Planck 
and TeV branes and calculable radiative corrections from gauge and 
matter loops.  Because the corrections of this order are still too 
large to satisfy constraints from the precision electroweak data, they 
have to be suppressed by a factor of a few to an order of magnitude. 
These may be attained by pushing up the mass scale of the resonances 
and/or by cancellations among various contributions.  It may also be 
possible for these corrections to mimic a light Higgs boson suggested 
by the precision electroweak data, by appropriately choosing parameters 
of the theory.  It will be interesting to study these corrections 
in more detail.

\section*{Acknowledgments}

I would like to thank David Smith for discussions.  I also thank 
Kaustubh Agashe and Alex Pomarol for useful discussions which 
have helped me to find and correct an error in the earlier version 
of this paper.  Fermilab is operated by Universities Research 
Association, Inc. for the U.S. Department of Energy under contract 
DE-AC02-76CH03000.


\newpage

\begin{thebibliography}{99}

\bibitem{Weinberg:gm}
S.~Weinberg,
Phys.\ Rev.\ D {\bf 13}, 974 (1976);
Phys.\ Rev.\ D {\bf 19}, 1277 (1979);
L.~Susskind,
Phys.\ Rev.\ D {\bf 20}, 2619 (1979).

\bibitem{Peskin:1990zt}
M.~E.~Peskin and T.~Takeuchi,
Phys.\ Rev.\ Lett.\  {\bf 65}, 964 (1990);
Phys.\ Rev.\ D {\bf 46}, 381 (1992).

\bibitem{Maldacena:1997re}
J.~M.~Maldacena,
Adv.\ Theor.\ Math.\ Phys.\  {\bf 2}, 231 (1998)
[Int.\ J.\ Theor.\ Phys.\  {\bf 38}, 1113 (1999)]
[arXiv:hep-th/9711200];
S.~S.~Gubser, I.~R.~Klebanov and A.~M.~Polyakov,
Phys.\ Lett.\ B {\bf 428}, 105 (1998)
[arXiv:hep-th/9802109];
E.~Witten,
Adv.\ Theor.\ Math.\ Phys.\  {\bf 2}, 253 (1998)
[arXiv:hep-th/9802150].

\bibitem{Arkani-Hamed:2000ds}
N.~Arkani-Hamed, M.~Porrati and L.~Randall,
JHEP {\bf 0108}, 017 (2001)
[arXiv:hep-th/0012148].

\bibitem{Randall:1999ee}
L.~Randall and R.~Sundrum,
Phys.\ Rev.\ Lett.\ {\bf 83}, 3370 (1999)
[arXiv:hep-ph/9905221].

\bibitem{Csaki:2003zu}
C.~Csaki, C.~Grojean, L.~Pilo and J.~Terning,
arXiv:hep-ph/0308038.

\bibitem{Agashe:2003zs}
K.~Agashe, A.~Delgado, M.~J.~May and R.~Sundrum,
arXiv:hep-ph/0308036.

\bibitem{Gherghetta:2003wm}
T.~Gherghetta and A.~Pomarol,
Phys.\ Rev.\ D {\bf 67}, 085018 (2003)
[arXiv:hep-ph/0302001].

\bibitem{Contino:2003ve}
R.~Contino, Y.~Nomura and A.~Pomarol,
arXiv:hep-ph/0306259.

\bibitem{Csaki:2003dt}
C.~Csaki, C.~Grojean, H.~Murayama, L.~Pilo and J.~Terning,
arXiv:hep-ph/0305237.

\bibitem{Nomura:2001mf}
Y.~Nomura, D.~R.~Smith and N.~Weiner,
Nucl.\ Phys.\ B {\bf 613}, 147 (2001)
[arXiv:hep-ph/0104041].

\bibitem{Grossman:1999ra}
Y.~Grossman and M.~Neubert,
Phys.\ Lett.\ B {\bf 474}, 361 (2000)
[arXiv:hep-ph/9912408].

\bibitem{Gherghetta:2000qt}
T.~Gherghetta and A.~Pomarol,
Nucl.\ Phys.\ B {\bf 586}, 141 (2000)
[arXiv:hep-ph/0003129].

\bibitem{Huber:2000ie}
S.~J.~Huber and Q.~Shafi,
Phys.\ Lett.\ B {\bf 498}, 256 (2001)
[arXiv:hep-ph/0010195].

\bibitem{Goldberger:2002cz}
W.~D.~Goldberger and I.~Z.~Rothstein,
Phys.\ Rev.\ Lett.\ {\bf 89}, 131601 (2002)
[arXiv:hep-th/0204160];
arXiv:hep-th/0208060.

\bibitem{Nomura:2003qb}
Y.~Nomura and D.~R.~Smith,
arXiv:hep-ph/0305214.

\bibitem{Davoudiasl:2000wi}
H.~Davoudiasl, J.~L.~Hewett and T.~G.~Rizzo,
Phys.\ Rev.\ D {\bf 63}, 075004 (2001)
[arXiv:hep-ph/0006041].

\bibitem{SekharChivukula:2001hz}
R.~Sekhar Chivukula, D.~A.~Dicus and H.~J.~He,
Phys.\ Lett.\ B {\bf 525}, 175 (2002)
[arXiv:hep-ph/0111016].

\end{thebibliography}
\end{document}